\documentclass[pre,twocolumn,preprintnumbers,amsmath,amssymb]{revtex4}
\usepackage{graphicx}% Include figure files
\usepackage{dcolumn}% Align table columns on decimal point
\usepackage{bm}% bold math

\begin{document}

\newcommand{\ie}{\textit{i.e. }}
\newcommand{\tmu}{\tilde{\mu}}
\newcommand{\TS}{\tilde{S}}
\newcommand{\TP}{\tilde{P}}
\newcommand{\TSTP}{\tilde{S}_{P}}
\newcommand{\beq}{\begin{equation}}
\newcommand{\eeq}{\end{equation}}
\newcommand{\beqn}{\begin{eqnarray}}
\newcommand{\eeqn}{\end{eqnarray}}
\newcommand{\omegat}{\bar{\omega}}

\title{Exclusion Zone of Convex Brushes in the Strong-Stretching Limit\footnote{The following article has been submitted to the Journal of Chemical Physics. After it is published, it will be found at http://ojps.aip.org/jcpo}}

\author{Vladimir A. Belyi}
 \email{v-belyi@uchicago.edu}
\affiliation{
James Franck Institute and the Department of Physics, University of Chicago \\
5640 S. Ellis Avenue, Chicago, Illinois 60637
}

\date{April 1, 2004}% It is always \today, today,

\begin{abstract}
We investigate asymptotic properties of long polymers grafted to convex cylindrical and spherical surfaces, and, in particular, distribution of chain free ends. The parabolic potential profile, predicted for flat and concave brushes, fails in convex brushes, and chain free ends span only a finite fraction of the brush thickness. In this paper, we extend the self-consistent model developed by Ball, Marko, Milner and Witten to determine the size of the exclusion zone, i.e. size of the region of the brush free from chain ends. We show that in the limit of strong stretching, the brush can be described by an alternative system of integral equations. This system can be solved exactly in the limit of weakly curved brushes, and numerically for the intermediate to strong curvatures. We find that going from melt state to theta solvent and then to marginal solvent decreases relative size of the exclusion zone. These relative differences grow exponentially as the curvature decreases to zero.
\end{abstract}

\pacs{82.35.Gh, 82.35.Lr, 83.80.Rs, 83.80.Sg}
\maketitle

\section{Introduction}

Long polymer chains grafted on a surface exhibit different behavior from their random-walk state. Limited space available to the attached chains leads to their strong extension. The first model of grafted chains was pioneered by Alexander \cite{Alexander1977} and de Gennes \cite{deGennes1976,deGennes1980,deGennes1985}, who assumed that all chain ends are localized on the outside of the brush. Later Semenov \cite{Semenov1985} and Milner, Witten, and Cates \cite{Milner1988a,Milner1988b,Milner1989} released the constraint on chain ends and developed a self-consistent field (SCF) theory of the brush. Similar approach was pursued by Zhulina and co-workers \cite{Zhulina1989a,Zhulina1989b,BirshteinZhulina1989}. The main conclusion of the SCF theory for flat brushes was that stretching of the chains is determined by a parabolic mean-field potential, and free chain ends span the whole area of the brush.

Similar results hold for concave brushes. However, in convex brushes, direct minimization of brush free energy in the SCF approach leads to unphysical state with negative density of chains. To explain this effect, Semenov \cite{Semenov1985} has proposed existence of an ``exclusion zone'', \ie region of a brush free of chain ends. Later Ball, Marko, Milner and Witten \cite{Ball1991} have developed an integral equation approach to a brush with exclusion zone that gave an exact solution for cylindrical brushes in a melt state. Subsequently Li and Witten \cite{LiWitten1994} have found exact solution for a cylindrical brush in marginal solvent in the limit of an infinite curvature.

The size of the exclusion zone in spherical brushes, as well as in cylindrical brushes in solvents at intermediate curvatures is still comparatively ill-understood. Several simulations \cite{Grest1994,Lindberg2001,Toral1993,MuratGrest1991,Carignano1995}, semi-analytical calculations \cite{WijmansZhulina1993,LiWitten1994}, and numerical approaches \cite{DanTirrell1992,WijmansZhulina1993} have been introduced. However, none of these fully exploits the limit of strong stretching to provide an analytical result at intermediate curvatures, or gives a quantitative answer in the limit of very strong curvatures.

In the present paper we investigate both cylindrical and spherical brushes in the melt, theta and marginal solvent states under the strong stretching assumption. We show that in this approximation brush properties may be described by a system of two integral equations. Compared with the results of Ball et all \cite{Ball1991}, proposed equations are linear and do not require self-consistent tuning of additional variables. In the limit of weak curvatures, these equation can be solved exactly for both melt and solvent states. We find that in this limit exclusion zone height exponentially decreases as radius of curvature increases. However, the rate of exponential decrease varies between solvent states, as well as between solvent and melt states. This suggests that the nature of the exclusion zone decrease is qualitatively different in various systems.

In the region of intermediate to strong curvatures, the brush equations can be solved numerically. We find that in the limit of strong curvature exclusion zone occupies between a minimum of 42.4\% of brush height for a cylindrical brush in a marginal solvent to a maximum of 75.9\% for a spherical brush in a melt. Finally, at all curvatures the relative size of the exclusion zone is larger in spherical than in cylindrical brushes. And in each of these systems exclusion zone decreases as one goes from melt to theta to marginal solvent.

\section{\label{sec:Model}Brushes Under Strong Stretching}

In the gaussian chain approximation, extension of a chain of $N$ elements to length $r$ costs $\frac{1}{2} (r^2/N a^2) (kT)$ in free energy. Here $a$ is the Kuhn's length of the chain segment, and $T$ is temperature. In the discussion that follows we will assume $kT=1$.

In grafted brushes, chains are additionally stretched by the repulsive forces from the adjacent chains, as monomers get expelled from the overfilled areas near the substrate. In the framework of a self-consistent field approach, effect of adjacent chains can be described by an effective potential $\mu(x)$. Physically this potential is related to the osmotic pressure acting at a given point in space and is determined by an average concentration of chain segments at that point. A given $\mu(x)$ thus determines the stretching profiles of the polymers and thence the monomer concentration profile. This profile must in turn be consistent with the original $\mu(x)$. Determining $\mu(x)$ is thus a problem of self-consistency.

The free energy of one chain in potential field $\mu(x)$ is given by

\begin{equation}
\label{eq_chainenergy}
F_c = \int_0^N dn \left[ \frac{1}{2 a^2} \left( \frac{dx}{dn} \right)^2 + \mu(x(n)) \right].
\end{equation}

The equilibrium properties of the brush are then determined by the partition function $Z = \sum \exp(-F[x(n)])$, where summation is carried over all possible chain conformations. For brushes with high grafting density, chains are strongly stretched so that the partition function $Z$ and other equilibrium properties are dominated by conformations with lowest free energy \cite{Semenov1985}. Hence, in the strongly-stretched limit all chain are assumed to be at their lowest energy state.

It was noticed by Semenov \cite{Semenov1985} and Milner et al \cite{Milner1988a,Milner1988b,Milner1989} that the free energy (\ref{eq_chainenergy}) has striking resemblance to the action of a classical particle moving in a potential field $-\mu(x)$. Hence the minimization problem is equivalent to finding a classical path of a particle moving in a given potential, with conservation of energy yielding

\begin{equation}
\label{eq:dxdn}
\frac{1}{2 a^2} \left( \frac{dx}{dn} \right)^2 = \mu(x) - \mu(x_0),
\end{equation}

\noindent
where constant of integration was chosen so that there is no pulling force on the open end of the chain: $(dx/dn)|_{x=x_0}=0$.

\begin{figure}[tb]
\centerline{\includegraphics[width=1.5in,height=1.5in] {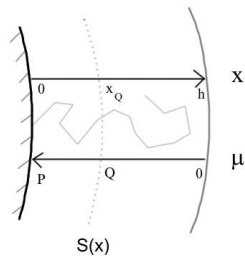}}
\caption{Geometry of the brush. $x_Q$ and $Q$ are coordinate and potential at the exclusion zone boundary. $S(x)$ is the area spanned by the chains at distance $x$ away from the grafting surface.}
\label{figCoordinates}
\end{figure}

In a mechanically stable brush, osmotic pressure is pushing chain segments away from the grafting surface, so that $\mu(x)$ is a monotonically decreasing function of $x$. Therefore, $\mu$ can be used in place of $x$ as a new coordinate (Fig. \ref{figCoordinates}).

Following eq. (\ref{eq:dxdn}), linear density of chain segments at point $\mu$ for the chains ending at point $\mu_0$ is given by

\beq
\left. \frac{dn}{dx} \right|_{\mu_0 \rightarrow \mu} = \left(\frac{1}{2a^2}\right)^{1/2} \frac{1}{[\mu - \mu_0]^{1/2}}.
\eeq

Now that we found chain extension in a given potential $\mu(x)$, we can proceed to the self-consistent constraints imposed on the potential. The first of these constraints is imposed by the geometrical restrictions on segment density. In melt, for instance, density fluctuations are prohibitively expensive, and local monomer concentration is assumed constant throughout the brush. In solvent, on the other hand, the concentration may vary, but it is directly related to the chemical potential via the free energy of mixing \cite{deGennesScalingConcepts}.

We start by introducing a function $S(x)$ related to the geometry of the brush. It describes the area available to chains at distance $x$ from the grafting surface (Fig. \ref{figCoordinates}), and, therefore, is presumed known. The physically most interesting geometries can be summarized by $S(x) = C_{d+1} (R + x)^d$, with $d$ being ``dimension'' of the grafting surface ($d=0, 1, 2$ for flat, cylindrical, and spherical surfaces respectively), and $R$ being the radius of curvature of the surface. The constant $C_{d} = d \cdot \pi^{d/2} / \Gamma(1 + d/2)$ is the area of a unit sphere in $d$ dimensions. Assuming that local volume fraction of monomers is $\phi(x)$, the total area occupied by monomers at distance $x$ is $\phi(x) S(x)$. In the melt state, the volume fraction is constant $\phi(x) = 1$, while in solvent state, $\phi$ varies with the distance. In the present paper, we will look at two types of solvents: a marginal solvent where $\mu \propto \phi$, and theta solvent where $\mu \propto \phi^2$ \cite{deGennesScalingConcepts}. Consideration of other solvents, such as general case of good solvents, would require detailed analysis of local monomer interactions in (\ref{eq_chainenergy}) and lies beyond the scope of this paper. Altogether $\phi(x) = \gamma_s \mu^s(x)$, where $s$ equals to $0, 1/2,$ and $1$ for melt, theta and marginal solvents respectively, and $\gamma_s$ is an unimportant proportionality coefficient.

The total area $\phi(x) S(x)$ spanned by the chains at distance $x$ from the grafting surface is comprised of the segments of the chains ending beyond $x$. Let $\sigma(x')$ be the total number of chains with free ends lying beyond $x'$. Then $\phi(x) S(x) = a_1^2 \int_x^h d\sigma(x') \left.\frac{dn}{dx}\right|_{x' \rightarrow x}$. Here $a_1$ is a microscopic length close to segment packing parameter. Using $\mu$ as a coordinate, the constraint on the segment density becomes

\beqn
\phi(\mu) S(\mu) & = & a_1^2 \int_0^\mu d\mu' \frac{d\sigma}{d\mu'}  \left.\frac{dn}{dx}\right|_{\mu' \rightarrow \mu} \nonumber \\
\label{eq_Svssigma}
 & = & \left(\frac{a_1^4}{2 a^2}\right)^{1/2} \int_0^\mu \frac{d\sigma}{d\mu'} \frac{d\mu'}{\left[\mu - \mu'\right]^{1/2}},
\eeqn

\noindent
where $S(\mu) \equiv S(x(\mu))$. The second line represents a well-known Abel equation \cite{TricomiIntegralEquations} and can be readily inverted

\beq
\label{eq_sigmavsS}
\sigma(\mu) = \left(\frac{2 a^2}{\pi^2 a_1^4}\right)^{1/2} \int_0^\mu \gamma_s \mu'^s S(\mu') \frac{d\mu'}{\left[\mu - \mu'\right]^{1/2}}.
\eeq

The second self-consistent constraint on the potential $\mu(x)$ is imposed by the length of the chains. The degree of polymerization for chains extended through point $\mu$ is given by

\beqn
N(\mu) & = & \left. \int_0^{x(\mu)} \frac{dn}{dx'}\right|_{\mu'\rightarrow\mu} dx' \nonumber \\
\label{eq_Nvsx}
& = & \left(\frac{1}{2 a^2}\right)^{1/2} \int_\mu^P \frac{dx}{d\mu'} \frac{d\mu'}{\left[\mu' - \mu\right]^{1/2}},
\eeqn

\noindent
where $P$ is potential at the grafting surface (Fig. \ref{figCoordinates}). Then condition of monodispersity would imply that $N(\mu) = N_0$ for all $\mu$ where chain ends are present.

In a simpler case of concave and flat brushes, the two integral equations (\ref{eq_sigmavsS}) and (\ref{eq_Nvsx}) can be readily solved. Indeed, in those brushes chain ends span the whole volume of the brush and $N(\mu) = N_0$ for all $\mu \in [0,P)$. Then equation (\ref{eq_Nvsx}) immediately yields parabolic profile $x \propto (P-\mu)^{1/2}$, so that chain distribution can be readily evaluated from (\ref{eq_sigmavsS}). However, in the case of convex brushes, this straightforward approach results in physically meaningless negative density of ends in the vicinity of the grafting surface. To avoid this unphysical behavior, one needs to introduce an ``exclusion zone'', \ie a region of the brush near the grafting surface with no free ends inside. We denote the potential at the boundary of this exclusion zone as $Q$. Then the constraint on monodispersity of chains reduces to $N(\mu) = N_0$ for $0 < \mu < Q$, while constraint on the absence of chain ends becomes $\sigma(\mu) = \sigma_0$ for $Q < \mu < P$.

Introduction of these constraints significantly complicates the problem, as two equations (\ref{eq_sigmavsS}) and (\ref{eq_Nvsx}) become interdependent. Still, in the melt state of cylindrical brushes, the two equation can be combined into single linear integral equation, that can later be solved exactly using electrostatic analogy \cite{Ball1991}. In the presence of solvent or spherical brushes, on the other hand, the final equation becomes more complicated, and even non-linear in the case of a spherical brush, with no exact solution known to the authors. Additional complications arise from the singular behavior of the unknown $dN/d\mu$ and $d\sigma/d\mu$.

To avoid this problem, we will rewrite equations (\ref{eq_sigmavsS}) and (\ref{eq_Nvsx}) in terms of one unknown function. The function $S(x)$ describing available volume as a function of distance from the surface is presumed to be known. The function $S(\mu)$, on the other hand, is not. Hence, if we reformulate and solve above equations in terms of this unknown function $S(\mu)$, then all the information about the system can be extracted from the comparison between $S(x)$ and $S(\mu) \equiv S(x(\mu))$.

We start with introduction of the dimensionless potentials $\tilde{\mu} = \mu / Q$ and $\tilde{P} = P / Q$, and proceed to rewriting equations (\ref{eq_Svssigma}) and (\ref{eq_sigmavsS}) in terms of $S(\tmu)$. Surface density of chains $\sigma(\tmu)$ is constant throughout the exclusion zone ($\tilde{\mu} > 1$), so that equation (\ref{eq_Svssigma}) can be rewritten as

\begin{multline}
\label{eq_Sabove}
S(\tmu > 1) = \frac{1}{\gamma_s \tmu^s Q^s} \left(\frac{a^2}{2 Q}\right)^{1/2} \int_0^1 \frac{d\sigma}{d\tmu'} \frac{d\tmu'}{\left[\tmu - \tmu'\right]^{1/2}}\\
= \frac{1}{\pi \tmu^s} \int_0^1 \tmu'^s S(\tmu') \frac{\left[\tmu - 1\right]^{1/2}}{\left[1-\tmu'\right]^{1/2} (\tmu - \tmu')} d\tmu',
\end{multline}

\noindent
where in the last step we substituted $d\sigma/d\mu'$ from (\ref{eq_sigmavsS}) and integrated out one of the integrals. The new equation allows to find $S(\tmu>1)$ if $S(\tmu<1)$ is known. Immediate advantage of this new integral equation is its implicit incorporation of the exclusion zone and independence of $Q$ or any other parameter of the system.

Similar procedure can be applied to (\ref{eq_Nvsx}). In this case $N(\tmu)$ is constant outside the exclusion zone ($\tmu < 1$), and $x(\mu) = C_d^{-1/d} S^{1/d}(\mu) - R$. Once again we can combine (\ref{eq_Nvsx}) and its inverse so that

\beq
\begin{split}
\label{eq_Sbelow}
\TS^{1/d}(\tmu & < 1) = \sqrt{1 - \tmu} + \alpha \; tan^{-1} \sqrt{\frac{1-\tmu}{\TP - 1}} \\
& + \frac{1}{\pi} \int_1^{\TP} \TS^{1/d}(\tmu') \frac{\left[1 - \tmu\right]^{1/2}}{\left[\tmu' - 1\right]^{1/2} (\tmu' - \tmu)} d\tmu'
\end{split}
\eeq

\noindent
where $\alpha = \left[R^2 / 2 N_0^2 Q a^2\right]^{1/2} = \frac{2}{\pi} \TS^{1/d}(\TP)$, and we have introduced rescaled $\TS(\tmu) = (\pi^2 / 8 N_0^2 Q a^2)^{d/2} S(\tmu)$. Obviously, this equation allows to find $\TS(\tmu < 1)$ if $\TS(\tmu > 1)$ is known. Due to linearity, similar rescaling of $S(\tmu)$ to $\TS(\tmu)$ can be performed in equation (\ref{eq_Sabove}), leading to a closed set of two integral equations (\ref{eq_Sabove}) and (\ref{eq_Sbelow}). Each one of these is linear, as opposed to the non-linear system of Ref \onlinecite{Ball1991}. Additionally, the system (\ref{eq_Sabove}, \ref{eq_Sbelow}) implicitly incorporates exclusion zone so that the structure of the final solutions is not hidden by additional constraints (such as requirement of positivity of end density \cite{Ball1991}).

The only parameter entering (\ref{eq_Sabove}) and (\ref{eq_Sbelow}) is the scaled potential $\TP$ which is related to the curvature of the grafting surface. In the melt state, this relation is immediately apparent from the conservation of brush volume. The volume of a curved brush is $C_{d+1} \left[ \left(R+h\right)^{d+1} - R^{d+1}\right]$, while the volume of a flat brush with the same number of chains and the same grafting density is $ (d+1) C_{d+1} R^d h_0$. Here $h_0$ is equilibrium thickness of the flat brush. Equating the two volumes we get

\begin{eqnarray}
\label{eq_Roverh0melt}
R/h_0 & = & \frac{d+1}{ \left(R+h\right)^{d+1} / R^{d+1} - 1 } \nonumber \\
& = & \frac{d+1}{\left[ \TS(\tmu=0 ) / \TS(\tmu=\TP)\right]^{1 + 1/d} - 1}.
\end{eqnarray}

In the presence of solvent, volume of the brush is no longer conserved and we need to use earlier definition of $\TS(x) = C_{d+1} (\pi^2 / 8 N_0^2 Q a^2)^{d/2} (R + x)^d$. Since grafting surface ($x=0$) corresponds to scaled potential $\tmu=\TP$, and open side of the brush ($x=h$) corresponds to $\tmu=0$, we may write

$$
R = \left(\frac{8 N_0^2 a^2 Q}{\pi^2 C_{d+1}^{2/d}} \right)^{1/2} \TS^{1/d}(\TP),
$$ 

\noindent 
and 
$$
h_0 = \left(\frac{8 N_0^2 a^2 Q_{flat}}{\pi^2 C_{d+1}^{2/d}} \right)^{1/2} [\TS^{1/d}_{flat}(1) - \TS^{1/d}_{flat}(0)],
$$

\noindent
where $S_{flat}$ refers to the flat brush, and $\TS^{1/d}_{flat}(1) - \TS^{1/d}_{flat}(0) = 1$, as it follows from the equation (\ref{eq_Sbelow}) in the limit of a flat brush ($\TP\rightarrow+1$). Hence $R/h_0 = (Q/Q_{flat})^{1/2} \TS^{1/d}(\TP)$. The ratio $Q/Q_{flat}$ of the exclusion zone potentials can be found from the equality of grafting densities in curved and flat brushes. This density is given by (\ref{eq_sigmavsS})

\beq
\frac{\sigma(\TP)}{S(\TP)} = 
\left(\frac{2 \gamma_s^2}{\pi^2 a^2}\right)^{1/2} Q^{s+1/2} 
\int_0^{\TP} \frac{\TS(\tmu')}{\TS(\TP)} \frac{ \tmu'^s d\tmu'}{[\TP - \tmu']^{1/2}}.
\eeq

In a flat brush $\TP \rightarrow 1$ and $\TS(\tmu) / \TS(\TP) = (R + x(\tmu)) / R \rightarrow 1$. Hence the radius of curvature and potential $\TP$ are related via

\beq
\label{eq_Roverh0}
R/h_0 = \TS^{1/d}(\TP) \left[ 
\frac{\Gamma(3/2+s)}{\pi^{1/2}\Gamma(1+s)} 
\int_0^{\TP} \frac{\TS(\tmu')}{\TS(\TP)} \frac{ \tmu'^s d\tmu'}{[\TP - \tmu']^{1/2}}
\right]^{-\frac{1}{2s+1}}.
\eeq

The remaining physical quantities can be immediately extracted from $\TS(\tmu)$. The exclusion zone $x_Q$, \ie the region of the brush near the grafting surface free of chain ends, can be found from $ (R+x_Q)^d / (R + h)^d = \TS(\tmu = 1) / \TS(\tmu = 0)$. Hence

\beq
\label{eq_xQ}
x_Q / h = \frac{\TS^{1/d}(\tmu=1) - \TS^{1/d}(\tmu=\TP)}{\TS^{1/d}(\tmu=0) - \TS^{1/d}(\tmu=\TP)}.
\eeq

\subsection{Weak Curvature Limit}

We now consider a limit of small curvatures. In this limit, exclusion zone height goes to zero, so that $\TP - 1 \ll 1$. Then function $\TS(\tmu)$ becomes nearly constant throughout the brush, and can be expanded about its value at the grafting surface $\TSTP = \TS(\TP)$: $\TS(\tmu>1) = \TSTP + \Delta \TS(\tmu)$ and $\TS(\tmu<1) = \TSTP + \Delta \TS(\tmu)$. Finally, it is convenient to introduce new variable $u = (1 - \tmu)^{1/2}$, for $\tmu < 1$, and $u = [(\tmu - 1) / (\TP - 1)]^{1/2}$ for $\tmu > 1$. 

To illustrate the approach, we start with the cylindrical melt brush ($s=0$ and $d=1$). Then equations (\ref{eq_Sabove}) and (\ref{eq_Sbelow}) transform into

\beq
\label{eq:TSA1}
\Delta \TS_>(u) = - \frac{2}{\pi} \TSTP \tan^{-1} (\epsilon_p u) + \frac{2}{\pi} \int_0^1 \Delta \TS_<(u') \frac{\epsilon_p u \; du'}{\epsilon_p^2 u^2 + u'^2},
\eeq

\noindent
and 

\beq
\label{eq:TSB1}
\Delta \TS_<(u) = u + \frac{2}{\pi} \int_0^1 \Delta \TS_>(u') \frac{\epsilon_p u \; du'}{u^2 + \epsilon_p^2 u'^2}.
\eeq

\noindent
Here $\epsilon_p = (\TP - 1)^{1/2}$. The last two equations are linear, and can be combined. Keeping only leading terms in $\epsilon_p$ we get

\beqn
\label{eq_cylFull}
\Delta \TS_>(u) & = &
  - \frac{2}{\pi} \TSTP \epsilon_p u 
  - \frac{2}{\pi} \epsilon_p u \ln(\epsilon_p u) \nonumber \\
  & & + \frac{4}{\pi^2} \int_0^1 \Delta \TS_>(u') \frac{u \ln(u/u') \; du'}{u^2 - u'^2}.
\eeqn

The curvature of the brush is related to the quantity $\epsilon_p$. Hence we are interested in the scaling behavior of $\TSTP$ and $\Delta \TS_>(u)$ with $\epsilon_p$. Applying standard separation of variables technique, we can regroup terms in (\ref{eq_cylFull}) based on their $u$ dependence and write down two separate equations for $\TSTP$ and $\Delta \TS_>(u)$:

\beq
\label{eq_cylA}
\TSTP + \ln(\epsilon_p) = \omegat,
\eeq

\beqn
\Delta \TS_>(u) = & - & \frac{2}{\pi} \omegat \epsilon_p u - \frac{2}{\pi} \epsilon_p u \ln(u) \nonumber \\
 & + & \frac{4}{\pi^2} \int_0^1 \Delta \TS_>(u') \frac{u \ln(\frac{u}{u'}) \; du'}{u^2 - u'^2}. \label{eq_cylB}
\eeqn

\noindent
Here $\omegat$ is a constant which, in general, may depend on $\epsilon_p$. The value of this constant is determined by the way original function $\TS(u)$ is split between $\TSTP$ and $\Delta \TS_>(u)$.  By construction, we want $\Delta \TS_>(u)$ to be exactly zero at $u=1$. Therefore substitution $\Delta \TS_>(u) = \frac{2}{\pi} \; G(u) \; \epsilon_p$ reduces Eq. (\ref{eq_cylB}) to

\beq
\label{eq_Gu}
G(u) = - \omegat u - u \ln(u) + \frac{4}{\pi^2} \int_0^1 G(u') \frac{u \ln(\frac{u}{u'}) \; du'}{u^2 - u'^2},
\eeq

\noindent
with additional constraint that $G(1)=0$. This equation has only one solution that satisfies given constraint. Using results of Ref. \onlinecite{Ball1991}, desired solution can be written in terms of the Meijer G-function \cite{MathMeijerGFunction,GradshteynRyzhik}:

\beq
\label{eq_GuSolution}
G(u)=\frac{\pi}{4} G_{2\;2}^{0\;2} \left( u^2 \left| 
\begin{array}{cc}
3/2 & 3/2\\
1 & 1
\end{array}
\right. \right).
\eeq

\noindent
The constant $\omegat$ introduced in (\ref{eq_cylA})-(\ref{eq_cylB}) is therefore also independent of $\epsilon_p$ and can be expressed in terms of the same Meijer G-function. Its numerical value is approximately $\omegat \approx 0.3864$. Hence the final solution for a cylindrical brush in a melt state is 

\beq
\TSTP = \omegat - \ln \epsilon_p,
\eeq

\noindent
and

\beq
\Delta \TS_>(u) = \frac{2}{\pi} \; G(u) \; \epsilon_p.
\eeq

As expected, when the brush becomes increasingly flat, $\TSTP \rightarrow \infty$, and $\Delta \TS_> \rightarrow 0$. Hence, the radius of curvature, given by Eqs. (\ref{eq_Roverh0melt}) and (\ref{eq_Roverh0}), and exclusion zone, Eq. (\ref{eq_xQ}), become

\beq
R/h_0 \approx - \frac{1}{2} + \TSTP^{1/d} = -\frac{1}{2} + \omegat - \ln \epsilon_p,
\eeq

\beq
x_Q/h \approx \Delta \TS_>(0) - \Delta \TS_>(1) = \frac{2}{\pi} \epsilon_p.
\eeq

\noindent
These are the desired results for the cylindrical brush in a melt state.

We now turn to the general case of a brush in a melt or solvent state. Once again, $\TS(\tmu)$ can be expanded about its value near the grafting surface $\TSTP$, so that to the leading order in $\epsilon_p$ equations (\ref{eq_Sabove}) and (\ref{eq_Sbelow}) reduce to

\beq
\label{eq_Sabove3}
\begin{split}
\Delta \TS_> & (u) =
  - \frac{2}{\pi} \TSTP  \left[ \frac{\Gamma(1+s)}{\Gamma(\frac{1}{2}+s)} \pi^{1/2} \right] \epsilon_p u  \\
  & + \frac{2}{\pi} \int_0^1 \Delta \TS_<(u') \left( \frac{1-u'^2}{1+\epsilon_p^2 u^2} \right)^s\frac{\epsilon_p u \; du'}{\epsilon_p^2 u^2 + u'^2},
\end{split}
\eeq

\noindent
and 

\beq
\label{eq_Sbelow3}
\Delta \TS_<(u) = 
  d \cdot \TSTP ^{(d-1)/d} u +
  \frac{2}{\pi} \int_0^1 \Delta \TS_>(u') \frac{\epsilon_p u \; du'}{u^2 + \epsilon_p^2 u'^2}.
\eeq

By analogy with the melt case discussed above, we substitute (\ref{eq_Sbelow3}) into (\ref{eq_Sabove3}) and break up the resulting equation into two separate equations for $\TSTP$ and $\Delta \TS_>(u)$:

\beq
\label{eq_generalA}
\frac{1}{d} \TSTP^{1/d} \frac{\Gamma(1+s)}{\Gamma(\frac{1}{2}+s)} \pi^{1/2} + \frac{1}{2} H_s + \ln(\epsilon_p) = \omegat,
\eeq

\beq
\label{eq_generalB}
\begin{split}
\Delta \TS_>(u) = & - \frac{2 d}{\pi} \TSTP^{(d-1)/d}\epsilon_p u \left[\omegat + \ln(u) \right] \\
& + \frac{4}{\pi^2} \int_0^1 \Delta \TS_>(u') \frac{u \ln(\frac{u}{u'}) \; du'}{u^2 - u'^2},
\end{split}
\eeq

\noindent
where $H_s$ is harmonic number \cite{MathHarmonicNumber}. These solve to

\beq
\TSTP^{1/d} = - \frac{d}{\pi^{1/2}} \frac{\Gamma(\frac{1}{2}+s)}{\Gamma(1+s)}  
\left[\frac{1}{2} H_s - \omegat + \ln \epsilon_p\right],
\eeq

\beq
\Delta \TS_>(u) = \frac{2 d}{\pi} \TSTP^{\frac{d-1}{d}}\; G(u) \; \epsilon_p.
\eeq

\noindent
Here $G(u)$ and $\omegat$ are the same variables as those introduced in equations (\ref{eq_Gu}) and (\ref{eq_GuSolution}). Once again, as the brush becomes increasingly flat, $\TSTP \rightarrow \infty$, and $\Delta \TS_> \rightarrow 0$. The radius of curvature (\ref{eq_Roverh0}) and exclusion zone (\ref{eq_xQ}) of a weakly curved brush are

\beqn
R/h_0 
& \approx & \TSTP^{1/d} - \frac{d}{\pi^{1/2}} \frac{\Gamma(\frac{3}{2}+s)}{\Gamma(2+s)}\\
&  = & - \frac{d}{\pi^{1/2}} \frac{\Gamma(\frac{1}{2}+s)}{\Gamma(1+s)}  
\left[ \frac{1}{2}H_s + \frac{1}{2 (s+1)} - \omegat + \ln \epsilon_p\right], \nonumber
\eeqn

\beq
x_Q/h \approx \frac{1}{d} \TSTP^{(1-d)/d} \left[\Delta \TS_>(0) - \Delta \TS_>(1) \right] = \frac{2}{\pi}  \epsilon_p.
\eeq

Finally, it is convenient to write down a relation between exclusion zone height and brush curvature. Using earlier convention where $d=1$ corresponds to the cylindrical brush, and $d=2$ corresponds to the spherical brush, the exclusion zone height in the weak curvature limit is given by

%\begin{gather}
%x_Q/h & \approx \frac{2}{\pi} e^{\left( \omegat - \frac{1}{2} \right) - \frac{1}{d} R/h_0}, \notag \\ 
%  \text{for melt} \label{eq_xQmelt} \\
%x_Q/h & \approx \frac{2}{\pi} e^{\left( \omegat - \frac{1}{3} - \frac{1}{2} H_{1/2} \right) - \frac{\pi}{2d} R/h_0}, \notag \\ 
%  \text{for theta solvent} \label{eq_xQTS} \\
%x_Q/h & \approx \frac{2}{\pi} e^{\left( \omegat - \frac{3}{4} \right) - \frac{2}{d} R/h_0}, \notag \\
%  \text{\label{eq_xQMS} for marginal solvent}
%\end{gather}

%\begin{align}
%x_Q/h & \approx \frac{2}{\pi} e^{\left( \omegat - \frac{1}{2} \right) - \frac{1}{d} R/h_0}, \notag \\
%  & \qquad \qquad \qquad \qquad \text{for melt}\label{eq_xQmelt} \\
%x_Q/h & \approx \frac{2}{\pi} e^{\left( \omegat - \frac{1}{3} - \frac{1}{2} H_{1/2} \right) - \frac{\pi}{2d} R/h_0}, \notag \\
%  & \qquad \qquad \qquad \qquad \text{for theta solvent} \label{eq_xQTS} \\
%x_Q/h & \approx \frac{2}{\pi} e^{\left( \omegat - \frac{3}{4} \right) - \frac{2}{d} R/h_0}, \notag \\
%  & \qquad \qquad \qquad \qquad \text{for marginal solvent} \label{eq_xQMS}
%\end{align}

%\begin{gather}
\begin{align}
\begin{split}
x_Q/h & \approx \frac{2}{\pi} e^{\left( \omegat - \frac{1}{2} \right) - \frac{1}{d} R/h_0}, \\
  & \qquad \qquad \qquad \qquad \qquad \text{for melt}
\end{split}\label{eq_xQmelt} \\
\begin{split}
x_Q/h & \approx \frac{2}{\pi} e^{\left( \omegat - \frac{1}{3} - \frac{1}{2} H_{1/2} \right) - \frac{\pi}{2d} R/h_0}, \\
  & \qquad \qquad \qquad \qquad \qquad \text{for theta solvent} 
\end{split} \label{eq_xQTS} \\
\begin{split}
x_Q/h & \approx \frac{2}{\pi} e^{\left( \omegat - \frac{3}{4} \right) - \frac{2}{d} R/h_0}, \\
  & \qquad \qquad \qquad \qquad \qquad \text{for marginal solvent} 
\end{split} \label{eq_xQMS}
\end{align}
%\end{gather}

\noindent
where harmonic number $H_{1/2} \approx 0.6137$ and constant $\omegat \approx 0.3864$.

\subsection{\label{sec:IntermediateCurvatures}Intermediate and Strong Curvatures}

For the general case of intermediate curvatures, we solve equations (\ref{eq_Sabove}) and (\ref{eq_Sbelow}) numerically. The unknown function $\TS(\tmu)$ is proportional to the area spanned by chains away from the grafting surface, and therefore is continuous and bounded everywhere on its domain. Hence we can approximate it by a piece-wise constant function and repeatedly iterate equations (\ref{eq_Sabove}) and (\ref{eq_Sbelow}) until a stable solution is achieved. It is known \cite{TricomiIntegralEquations} that iteration procedure for linear Fredholm equations with $L_2$ kernels converges when eigenvalue is small compared to the norm of kernel. Furthermore, Fredholm equations are also solvable when kernels have finite number of simple poles. While these may not be used as a proof of convergences of the non-linear system (\ref{eq_Sabove}, \ref{eq_Sbelow}), the convergence may reasonably be expected. To further justify it, we have tried various initial states to verify that the system always iterates to the same final state.

\begin{figure}[tb]
\centerline{\includegraphics[width=3.4in,height=2.4in] {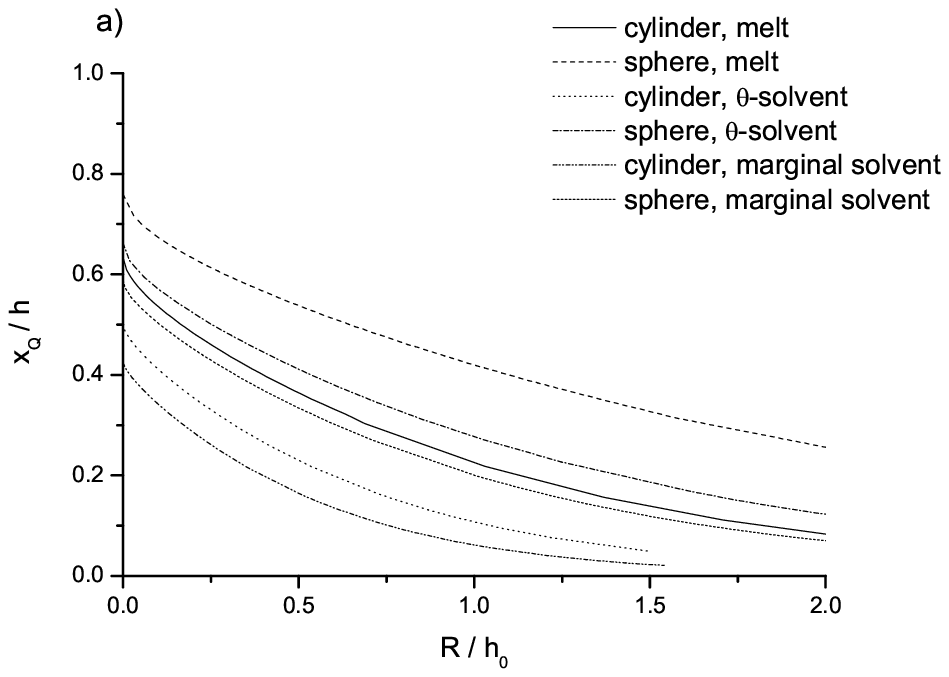}}
\centerline{\includegraphics[width=3.4in,height=2.4in] {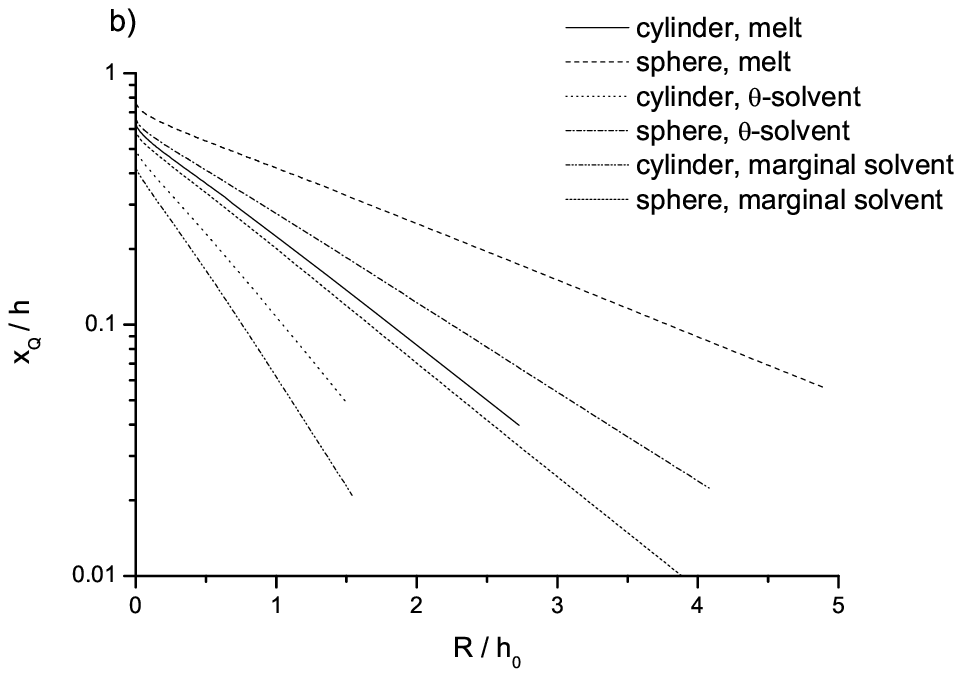}}
\caption{Exclusion zone heights for cylindrical and spherical brushes in melt and solvent states for a) strongly curved brushes, and b) weakly curved brushes.}
\label{figxQChart}
\end{figure}

In the present work the function domain was split into 4,000 segments, and iterations were repeated until accuracy of $10^{-9}$ was achieved. Use of piece-wise constant functions made it possible to analytically integrate out all the singularities, so that no further treatment of singular points was necessary. The results of numerical iterations are summarized in Figures \ref{figxQChart}.

\section{Discussion}

We have shown that exclusion zone is characteristic property of a brush in all considered solvent and melt states, for both cylindrical and spherical brushes. The size of an exclusion zone monotonically increases with curvature, until it reaches its finite maximum value at the extreme curvature limit. Our results are in perfect agreement with the known exact results for certain cylindrical brushes \cite{Ball1991,LiWitten1994}. Namely, we observe exact match with the exclusion zone height in the cylindrical melt calculated for all curvatures by Ball et al \cite{Ball1991}. Also, exclusion zone height matches with the exact result for a strongly curved cylindrical brush in marginal solvent calculated by Li and Witten \cite{LiWitten1994}. However, it should come as no surprise, as all of these methods are based on similar SCF approach.

\begin{table}  
  \caption{\label{tableExtremeCurvatures}Relative size of the exclusion zone ($x_Q/h$) at the limit of strong curvature for different brushes.}
  \begin{ruledtabular}
    \begin{tabular}{lcc}
      & Cylinder & Sphere \\
      \hline\\
      melt & 0.6340 & 0.7594 \\
      theta solvent & 0.4999 & 0.6621 \\
      marginal solvent & 0.4242 & 0.5872 \\
    \end{tabular} 
  \end{ruledtabular}
\end{table}

\begin{table}  
  \caption{\label{tableWeakCurvatureVSNumerics}Coefficients in the exponential dependence $A e^{-\alpha R/h_0}$ of the exclusion zone height on radius of curvature. Predicted values are based on the weak curvature expansion, Eqs. (\ref{eq_xQmelt}) - (\ref{eq_xQMS}), and best-fit values are from numerical calculations, Fig. \ref{figxQChart}}
  \begin{ruledtabular}
    \begin{tabular}{lcccc}
      &\multicolumn{2}{c}{Best-fit}&\multicolumn{2}{c}{Predicted}\\
      & $A$ & $\alpha$ & $A$ & $\alpha$ \\
      \hline\\
      cylindrical brush, melt state & 0.63 & 0.99 & 0.57 & 1 \\
      cylindrical brush in theta solvent & 0.52 & 1.56 & 0.49 & $\pi/2$ \\
      cylindrical brush in marginal solvent & 0.46 & 2.00 & 0.44 & 2 \\
      spherical brush, melt state & 0.71 & 0.52 & 0.57 & 1/2 \\
      spherical brush in theta solvent & 0.60 & 0.81 & 0.49 & $\pi/4$ \\
      spherical brush in marginal solvent & 0.52 & 1.01 & 0.44 &  1 \\
    \end{tabular} 
  \end{ruledtabular}		
\end{table}

Exact results for the weak curvature limit are summarized by equations (\ref{eq_xQmelt}) - (\ref{eq_xQMS}). As one may expect, at weak curvatures dimension of the brush (i.e cylinder vs. sphere) is insignificant and exclusion zone height varies solely with the mean radius of curvature $R/d$. However, the strength of curvature dependence changes with solvent properties: decrease of the exclusion zone with radius is fastest for marginal solvent, and is slowest for the melt. This might explain why it was not observed in earlier solvent simulations by Murat and Grest \cite{MuratGrest1991}, even though later simulations by Grest \cite{Grest1994} did indicate existence of finite exclusion zone. Besides falling off faster, we also predict exclusion zone at extreme curvatures to be smallest for marginal solvents, and largest for the melt, with theta solvent in between (Table \ref{tableExtremeCurvatures}). Comparison between analytical results in the weak curvature limit, Eqs. (\ref{eq_xQmelt}) - (\ref{eq_xQMS}) and results of numerical calculations are summarized in Table \ref{tableWeakCurvatureVSNumerics}. These results indicate very good agreement in the slope of the exponential decay. However, weak curvature limit, i.e. limit of $R \rightarrow \infty$, gives somewhat underestimated values of the intercept point in the depicted range of curvatures.

In a spherical brush under marginal solvent conditions, the maximum size of the exclusion zone is $x_Q \approx 0.5872\;h$. This is significantly lower than 94\% estimated by Li and Witten \cite{LiWitten1994} using variational approach. However, such a difference is not surprising, as these authors have found that exact solution for the cylindrical brush is also significantly lower than the value predicted from variational approach. Additionally, lower size of the exclusion zone predicted in this work is in line with later simulations by Grest \cite{Grest1994} and numerical calculations by Wijmans and Zhulina \cite{WijmansZhulina1993}, and Dan and Tirrell \cite{DanTirrell1992}.

\begin{figure}[tb]
\centerline{\includegraphics[width=3.2in,height=2.2in] {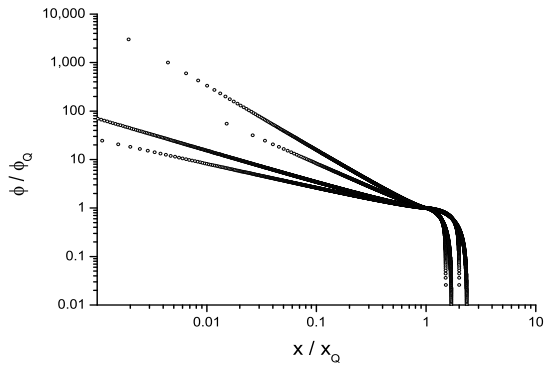}}
\caption{Volume fraction of chain segments in the case of extreme curvature ($R=0$). All values are normalized to one at the boundary of the exclusion zone. In the order of increasing slope, the curves correspond to a) cylindrical brush in a theta solvent, b) cylindrical brush in a marginal solvent, c) spherical brush in a theta solvent, and d) spherical brush in a marginal solvent.}
\label{fig:SegmentDensity}
\end{figure}

We find that the size of exclusion zone is always a monotonically increasing function of curvature, similar to the cylindrical brushes under melt conditions \cite{Ball1991}. However, this contradicts with numerical calculations performed by Dan and Tirrell, who used diffusion-equation approach to the self-consistent problem \cite{DanTirrell1992}. These authors observed a sharp decrease in exclusion zone size at very strong curvatures, and it was found to be completely independent of any parameters of the system. However, these authors also observed unexpected drop in monomer density near the grafting surface. Hence one may speculate that reported decrease in exclusion zone height is not a property of the diffusion-equation approach, but is rather a consequence of diverging numerical calculations near singularity. 

Still, possibility of a decrease in the exclusion zone height at large curvatures remains open. Stretching of the chains continuously decreases away from the grafting surface, and fluctuations should play larger role in the outer region of the brush. As this effect becomes even more pronounced in strongly curved brushes, it may affect exclusion zone height and cause its decrease at large curvatures. We will leave this question for future research.

The volume fraction $\phi(x)$ of chain segments in different brushes in the limit of extreme curvature is shown in Figure \ref{fig:SegmentDensity}. Deep inside the exclusion zone ($x \ll x_Q$) the volume fraction follows a power law, with corresponding best-fit exponents shown in Table \ref{table:PhiScaling}. These fitting exponents are in excellent agreement with predictions of a largely simplified model of the brush, which assumes that all chain ends lie on the outside of the brush. That model, first pioneered by Alexander \cite{Alexander1977} and de Gennes \cite{deGennes1976,deGennes1980,deGennes1985} and later extended by Daoud and Cotton \cite{DaoudCotton1982} for spherical brushes, and Birshtein and Zhulina \cite{BirshteinZhulina1984} for cylindrical brushes, predicts that the volume fraction of chain segments should scale with distance from the grafting surface as

\beq
\label{eq:Daoud}
\phi(x) \propto (R + x)^{-d \cdot (2\nu - 1) / 2\nu},
\eeq

\noindent
where $d$ is dimensionality of the brush ($d=1$ for cylindrical brush, and $d=2$ for spherical brush), and the Flory exponent $\nu$ depends on the solvent: $\nu=3/5$ for good solvent, and $\nu=1/2$ for theta solvent \cite{WijmansZhulina1993}. In the vicinity of the grafting surface, \ie far away from the exclusion zone boundary, results of our model match with Eq. (\ref{eq:Daoud}), as shown in Table \ref{table:PhiScaling}. However, as one approaches the boundary of the exclusion zone, deviations from scaling models arise, and the power law exponent becomes smaller than that predicted by (\ref{eq:Daoud}). Outside exclusion zone, the density profile falls off much faster and no longer follows the power law.

\begin{table}  
  \caption{\label{table:PhiScaling}Power law exponents for the dependence of volume fraction on distance from grafting surface $\phi(x) \propto x^\alpha$. Predicted values are based on equation (\ref{eq:Daoud}). }
  \begin{ruledtabular}
    \begin{tabular}{lcc}
      & Best-fit & Predicted \\
      \hline\\
      cylindrical brush in theta solvent & -0.4932 & - 1/2 \\
      cylindrical brush in marginal solvent & -0.6660 & - 2/3 \\
      spherical brush in theta solvent & -0.9871 & - 1 \\
      spherical brush in marginal solvent & -1.332 & - 4/3 \\
    \end{tabular} 
  \end{ruledtabular}		
\end{table}

Similarly, in the brushes with non-zero radius of curvature $R$, simplified model of Daoud and Cotton becomes inapplicable and deviations from density profile (\ref{eq:Daoud}) increase. Qualitatively, absolute value of the power law exponent decreases as radius increases.

\begin{figure}[tb]
\centerline{\includegraphics[width=3.2in,height=2.2in] {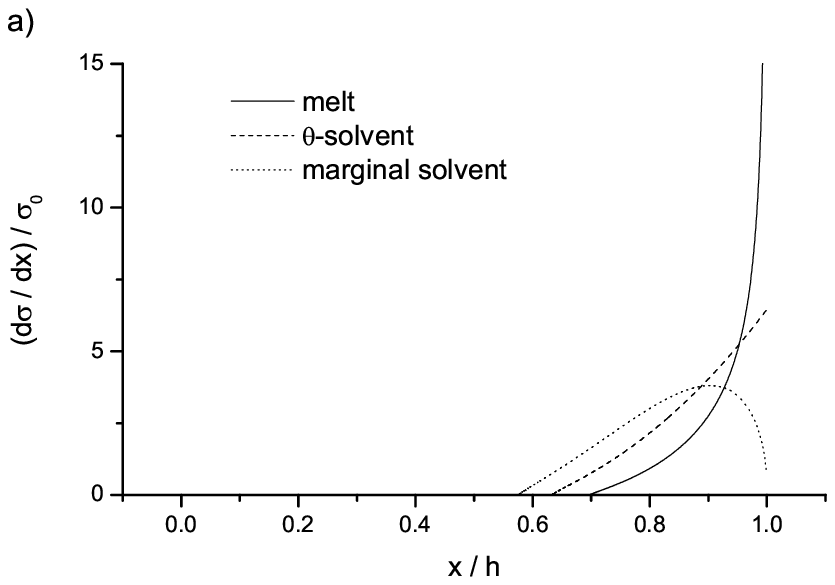}}
\centerline{\includegraphics[width=3.2in,height=2.2in] {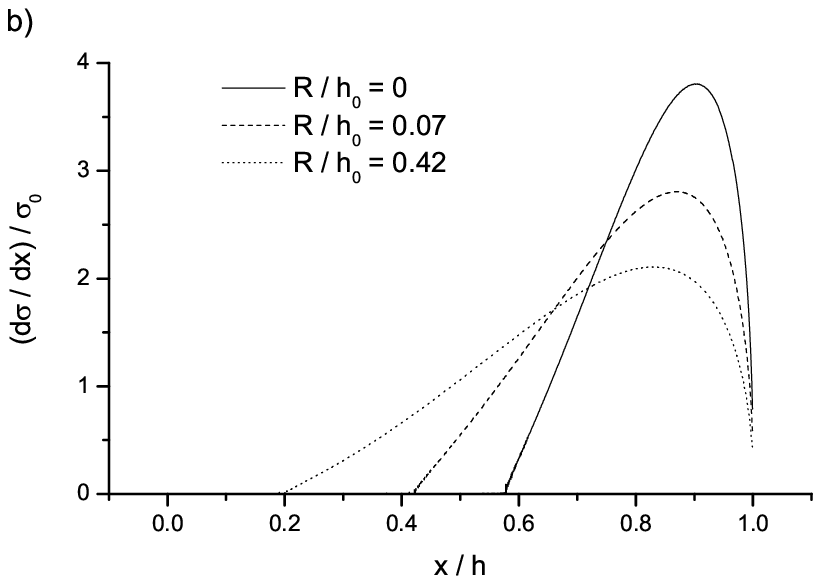}}
\caption{Density of free ends in a spherical brush a) at strong curvature limit, and b) in marginal solvent at different curvatures. }
\label{fig:EndDensity}
\end{figure}

Finally, the density profiles for the chain free ends are shown on Figure \ref{fig:EndDensity}. The presence of the exclusion zone does restrict chain ends to a fraction of a brush, but it does not qualitatively affect their distribution: at the boundary of the exclusion zone the density of ends continuously approaches zero, while at the open surface free ends distribution maintains the same structure as found in flat and concave brushes.

\section{Conclusions}

In conclusion, we have shown that exclusion zone occupies a finite fraction of the brush in all convex geometries considered. In both cylindrical and spherical brushes, in either solvent or melt state, the size of the exclusion zone is a monotonic function of curvature, continuously increasing as radius of curvature goes to zero. In the limit of strong curvature it reaches the maximum value, which is always a finite fraction of the brush thickness (42\% to 76\% of the brush, Table \ref{tableExtremeCurvatures}).

At weak curvatures exclusion zone height decreases exponentially with radius. However, rate of this decrease depends on the solvent/melt state of the system, indicating qualitatively different nature of the underlying mechanism.

In the vicinity of the grafting surface the properties of the brush are very well described by the scaling model of Daoud and Cotton \cite{DaoudCotton1982}. The match becomes nearly perfect in the limit of strong curvatures (Table \ref{table:PhiScaling}). However, away from the grafting surface, as one approaches boundary of the exclusion zone, the deviations arise. And, ultimately, at the opposite side of the brush, i.e. near the open surface, the brush properties approach those predicted by Semenov \cite{Semenov1985} and Milner et al \cite{Milner1988a,Milner1988b,Milner1989}.

Analysis given in appendix shows that approximation of strong stretching is suitable for most brushes in the limit of large degree of polymerization. Hence it should provide a good qualitative guide to the real systems, and in particular to the colloidal particles or comb polymers in a solvent environment. Detailed knowledge of chain conformations in these brushes may provide new mechanisms of controlling properties of these systems.
 
Finally, the approach developed in this paper may prove useful in analyzing brushes with inhomogeneous structures, and, in particular, brushes with polydisperse chains or multi-component systems. Investigation of chain conformations in these systems, as well as analysis of fluctuations due to non-asymptotic stretching, are some of the possible areas of future research.

\section*{Acknowledgments}

This research was performed in partial fulfillment of the doctorate degree requirements in physics at the University of Chicago. I am particularly grateful to my advisor, Prof. Thomas Witten, for both suggesting the problem and providing ongoing supervision of my research. This work was supported in part by the MRSEC Program of the National Science Foundation under Award Number DMR-0213745. 

\appendix* 

\section{Applicability of the strong stretching approximation.}

The approximation used in this paper was based on two assumptions. First of all, grafted chains were assumed to be significantly stretched compared to their free radius of gyration. This allowed us to neglect fluctuations in chain trajectories. On the other hand, chain extension was assumed to be significantly smaller than the full extended length to ensure validity of the gaussian approximation (\ref{eq_chainenergy}). Introduction of these constraints imposes some limits on the applicability of the theory, which are discussed below.

Our analysis here is similar to that of Ref. \onlinecite{Ball1991}, so we will briefly summarize the main conclusions and extend them to the broader class of solvent states considered in this paper. 

The length of a chain in a dense brush is of the same order as the brush thickness $h$. Therefore, aforementioned constraints on chain length can be written as

\beq
\label{eq:approximation}
a N^{1/2} \ll h \ll a N,
\eeq

\noindent
where $N$ is chain polymerization, and $a$ is characteristic size of one monomer. Here we are interested in asymptotic behavior at large $N$.

Let us first consider flat and weakly curved brushes. In a melt state flat brush has a thickness $h \propto N a^3 \sigma / A$, where $\sigma$ is the total number of grafted chains, and $A$ is total area of the grafting surface. Therefore, constraints (\ref{eq:approximation}) are asymptotically satisfied by brushes with low grafting density. In the solvent state, brush height can be estimated from minimization of average free energy per chain $F = h^2 / 2 N a^2 + w_\nu \phi^\nu V / \sigma$, where $\nu$ describes solvent state ($\nu=2$ for marginal solvent, and $\nu=3$ for theta solvent), $V = A h$ is the total volume of the brush, and $w_\nu$ is unimportant proportionality coefficient related to the virial coefficients. The volume fraction of monomers in a flat brush is $\phi = \sigma a^3 N / V$, and minimization of free energy gives $h \propto (\sigma / A)^{(\nu - 1) / (\nu + 1)} N$. Similarly to the melt case, nearly flat brushes in a solvent also asymptotically satisfy strong stretching approximation at low grafting density.

\begin{table}
  \caption{\label{tableScalingLaws}Scaling of brush height with grafting density and chain polymerization for strongly curved brushes.}
  \begin{ruledtabular}
    \begin{tabular}{lcc}
      & cylinder & sphere \\
      \hline\\
      melt & $\sigma^{1/2} N^{1/2}$ & $\sigma^{1/3} N^{1/3}$ \\
      theta solvent & $\sigma^{1/3} N^{2/3}$ & $\sigma^{1/4} N^{1/2}$ \\
      marginal solvent & $\sigma^{1/4} N^{3/4}$ & $\sigma^{1/5} N^{3/5}$ \\
    \end{tabular} 
  \end{ruledtabular}		
\end{table}

Similar considerations may be applied to the curved brushes. Considering the limit of strong curvatures, the scaling laws for the brush heights take the form shown in Table \ref{tableScalingLaws}. Evidently, most brushes once again satisfy Eq. (\ref{eq:approximation}): the cylindrical and spherical brushes in marginal solvent, as well as cylindrical brushes in theta solvent always obey (\ref{eq:approximation}) in the asymptotic limit of large $N$. The cylindrical brushes in a melt state, and spherical brushes in theta solvent asymptotically satisfy (\ref{eq:approximation}), provided that their grafting densities are sufficiently large. Here large densities mean that the total number of chains $\sigma \gg w_3^{1/8} a^{11/8}$ for the sphere, and linear density of chains $\sigma \gg a^{-1}$ for the cylinder.

The only system that violates (\ref{eq:approximation}) in the asymptotic limit of large $N$ is the spherical brush in a melt state. Specifically, this system fails the first of the two constraints, and therefore requires detailed accounting of different chain trajectories. Still, our theory should correctly describe this brush in the limit of weak curvatures, and may give qualitative insights for the case of intermediate to strong curvatures.

\bibliography{Citations}

\begin{thebibliography}{27}
\expandafter\ifx\csname natexlab\endcsname\relax\def\natexlab#1{#1}\fi
\expandafter\ifx\csname bibnamefont\endcsname\relax
  \def\bibnamefont#1{#1}\fi
\expandafter\ifx\csname bibfnamefont\endcsname\relax
  \def\bibfnamefont#1{#1}\fi
\expandafter\ifx\csname citenamefont\endcsname\relax
  \def\citenamefont#1{#1}\fi
\expandafter\ifx\csname url\endcsname\relax
  \def\url#1{\texttt{#1}}\fi
\expandafter\ifx\csname urlprefix\endcsname\relax\def\urlprefix{URL }\fi
\providecommand{\bibinfo}[2]{#2}
\providecommand{\eprint}[2][]{\url{#2}}

\bibitem[{\citenamefont{Alexander}(1977)}]{Alexander1977}
\bibinfo{author}{\bibfnamefont{S.~J.} \bibnamefont{Alexander}},
  \bibinfo{journal}{J. Phys. (Paris)} \textbf{\bibinfo{volume}{38}},
  \bibinfo{pages}{983} (\bibinfo{year}{1977}).

\bibitem[{\citenamefont{de~Gennes}(1976)}]{deGennes1976}
\bibinfo{author}{\bibfnamefont{P.-G.} \bibnamefont{de~Gennes}},
  \bibinfo{journal}{J. Phys. (Paris)} \textbf{\bibinfo{volume}{37}},
  \bibinfo{pages}{1443} (\bibinfo{year}{1976}).

\bibitem[{\citenamefont{de~Gennes}(1980)}]{deGennes1980}
\bibinfo{author}{\bibfnamefont{P.-G.} \bibnamefont{de~Gennes}},
  \bibinfo{journal}{Macromolecules} \textbf{\bibinfo{volume}{13}},
  \bibinfo{pages}{1069} (\bibinfo{year}{1980}).

\bibitem[{\citenamefont{de~Gennes}(1985)}]{deGennes1985}
\bibinfo{author}{\bibfnamefont{P.-G.} \bibnamefont{de~Gennes}},
  \bibinfo{journal}{C. R. Acad. Sci. (Paris)} \textbf{\bibinfo{volume}{300}},
  \bibinfo{pages}{839} (\bibinfo{year}{1985}).

\bibitem[{\citenamefont{Semenov}(1985)}]{Semenov1985}
\bibinfo{author}{\bibfnamefont{A.~N.} \bibnamefont{Semenov}},
  \bibinfo{journal}{Sov. Phys. JETP} \textbf{\bibinfo{volume}{61}},
  \bibinfo{pages}{733} (\bibinfo{year}{1985}).

\bibitem[{\citenamefont{Milner et~al.}(1988{\natexlab{a}})\citenamefont{Milner,
  Witten, and Cates}}]{Milner1988a}
\bibinfo{author}{\bibfnamefont{S.~T.} \bibnamefont{Milner}},
  \bibinfo{author}{\bibfnamefont{T.~A.} \bibnamefont{Witten}},
  \bibnamefont{and} \bibinfo{author}{\bibfnamefont{M.~E.} \bibnamefont{Cates}},
  \bibinfo{journal}{Macromolecules} \textbf{\bibinfo{volume}{21}},
  \bibinfo{pages}{2610} (\bibinfo{year}{1988}{\natexlab{a}}).

\bibitem[{\citenamefont{Milner et~al.}(1988{\natexlab{b}})\citenamefont{Milner,
  Witten, and Cates}}]{Milner1988b}
\bibinfo{author}{\bibfnamefont{S.~T.} \bibnamefont{Milner}},
  \bibinfo{author}{\bibfnamefont{T.~A.} \bibnamefont{Witten}},
  \bibnamefont{and} \bibinfo{author}{\bibfnamefont{M.~E.} \bibnamefont{Cates}},
  \bibinfo{journal}{Europhys. Lett.} \textbf{\bibinfo{volume}{5}},
  \bibinfo{pages}{413} (\bibinfo{year}{1988}{\natexlab{b}}).

\bibitem[{\citenamefont{Milner et~al.}(1989)\citenamefont{Milner, Witten, and
  Cates}}]{Milner1989}
\bibinfo{author}{\bibfnamefont{S.~T.} \bibnamefont{Milner}},
  \bibinfo{author}{\bibfnamefont{T.~A.} \bibnamefont{Witten}},
  \bibnamefont{and} \bibinfo{author}{\bibfnamefont{M.~E.} \bibnamefont{Cates}},
  \bibinfo{journal}{Macromolecules} \textbf{\bibinfo{volume}{22}},
  \bibinfo{pages}{853} (\bibinfo{year}{1989}).

\bibitem[{\citenamefont{Zhulina et~al.}(1989)\citenamefont{Zhulina, Pryamitsyn,
  and Borisov}}]{Zhulina1989a}
\bibinfo{author}{\bibfnamefont{E.~B.} \bibnamefont{Zhulina}},
  \bibinfo{author}{\bibfnamefont{V.~A.} \bibnamefont{Pryamitsyn}},
  \bibnamefont{and} \bibinfo{author}{\bibfnamefont{O.~V.}
  \bibnamefont{Borisov}}, \bibinfo{journal}{Vysokomol. Soedin. Ser. A}
  \textbf{\bibinfo{volume}{31}}, \bibinfo{pages}{185} (\bibinfo{year}{1989}).

\bibitem[{\citenamefont{Zhulina and Semenov}(1989)}]{Zhulina1989b}
\bibinfo{author}{\bibfnamefont{E.~B.} \bibnamefont{Zhulina}} \bibnamefont{and}
  \bibinfo{author}{\bibfnamefont{A.~N.} \bibnamefont{Semenov}},
  \bibinfo{journal}{Vysokomol. Soedin. Ser. A} \textbf{\bibinfo{volume}{31}},
  \bibinfo{pages}{177} (\bibinfo{year}{1989}).

\bibitem[{\citenamefont{Birshtein and Zhulina}(1989)}]{BirshteinZhulina1989}
\bibinfo{author}{\bibfnamefont{T.~M.} \bibnamefont{Birshtein}}
  \bibnamefont{and} \bibinfo{author}{\bibfnamefont{E.~B.}
  \bibnamefont{Zhulina}}, \bibinfo{journal}{Polymer}
  \textbf{\bibinfo{volume}{30}}, \bibinfo{pages}{170} (\bibinfo{year}{1989}).

\bibitem[{\citenamefont{Ball et~al.}(1991)\citenamefont{Ball, Marko, Milner,
  and Witten}}]{Ball1991}
\bibinfo{author}{\bibfnamefont{R.~C.} \bibnamefont{Ball}},
  \bibinfo{author}{\bibfnamefont{J.~F.} \bibnamefont{Marko}},
  \bibinfo{author}{\bibfnamefont{S.~T.} \bibnamefont{Milner}},
  \bibnamefont{and} \bibinfo{author}{\bibfnamefont{T.~A.}
  \bibnamefont{Witten}}, \bibinfo{journal}{Macromolecules}
  \textbf{\bibinfo{volume}{24}}, \bibinfo{pages}{693} (\bibinfo{year}{1991}).

\bibitem[{\citenamefont{Li and Witten}(1994)}]{LiWitten1994}
\bibinfo{author}{\bibfnamefont{H.}~\bibnamefont{Li}} \bibnamefont{and}
  \bibinfo{author}{\bibfnamefont{T.~A.} \bibnamefont{Witten}},
  \bibinfo{journal}{Macromolecules} \textbf{\bibinfo{volume}{27}},
  \bibinfo{pages}{449} (\bibinfo{year}{1994}).

\bibitem[{\citenamefont{Grest}(1994)}]{Grest1994}
\bibinfo{author}{\bibfnamefont{G.~S.} \bibnamefont{Grest}},
  \bibinfo{journal}{Macromolecules} \textbf{\bibinfo{volume}{27}},
  \bibinfo{pages}{3493} (\bibinfo{year}{1994}).

\bibitem[{\citenamefont{Lindberg and Elvingson}(2001)}]{Lindberg2001}
\bibinfo{author}{\bibfnamefont{E.}~\bibnamefont{Lindberg}} \bibnamefont{and}
  \bibinfo{author}{\bibfnamefont{C.}~\bibnamefont{Elvingson}},
  \bibinfo{journal}{J. Chem. Phys.} \textbf{\bibinfo{volume}{114}},
  \bibinfo{pages}{6343} (\bibinfo{year}{2001}).

\bibitem[{\citenamefont{Toral and Chakrabarti}(1993)}]{Toral1993}
\bibinfo{author}{\bibfnamefont{R.}~\bibnamefont{Toral}} \bibnamefont{and}
  \bibinfo{author}{\bibfnamefont{A.}~\bibnamefont{Chakrabarti}},
  \bibinfo{journal}{Phys. Rev. E} \textbf{\bibinfo{volume}{47}},
  \bibinfo{pages}{4240} (\bibinfo{year}{1993}).

\bibitem[{\citenamefont{Murat and Grest}(1991)}]{MuratGrest1991}
\bibinfo{author}{\bibfnamefont{M.}~\bibnamefont{Murat}} \bibnamefont{and}
  \bibinfo{author}{\bibfnamefont{S.}~\bibnamefont{Grest}},
  \bibinfo{journal}{Macromolecules} \textbf{\bibinfo{volume}{24}},
  \bibinfo{pages}{704} (\bibinfo{year}{1991}).

\bibitem[{\citenamefont{Carignano and Szleifer}(1995)}]{Carignano1995}
\bibinfo{author}{\bibfnamefont{M.~A.} \bibnamefont{Carignano}}
  \bibnamefont{and} \bibinfo{author}{\bibfnamefont{I.}~\bibnamefont{Szleifer}},
  \bibinfo{journal}{J. Chem. Phys.} \textbf{\bibinfo{volume}{102}},
  \bibinfo{pages}{8662} (\bibinfo{year}{1995}).

\bibitem[{\citenamefont{Wijmans and Zhulina}(1993)}]{WijmansZhulina1993}
\bibinfo{author}{\bibfnamefont{C.~M.} \bibnamefont{Wijmans}} \bibnamefont{and}
  \bibinfo{author}{\bibfnamefont{E.~B.} \bibnamefont{Zhulina}},
  \bibinfo{journal}{Macromolecules} \textbf{\bibinfo{volume}{26}},
  \bibinfo{pages}{7214} (\bibinfo{year}{1993}).

\bibitem[{\citenamefont{Dan and Tirrell}(1992)}]{DanTirrell1992}
\bibinfo{author}{\bibfnamefont{N.}~\bibnamefont{Dan}} \bibnamefont{and}
  \bibinfo{author}{\bibfnamefont{M.}~\bibnamefont{Tirrell}},
  \bibinfo{journal}{Macromolecules} \textbf{\bibinfo{volume}{25}},
  \bibinfo{pages}{2890} (\bibinfo{year}{1992}).

\bibitem[{\citenamefont{de~Gennes}(1979)}]{deGennesScalingConcepts}
\bibinfo{author}{\bibfnamefont{P.-G.} \bibnamefont{de~Gennes}},
  \emph{\bibinfo{title}{Scaling Concepts in Polymer Physics}}
  (\bibinfo{publisher}{Cornell University Press: Ithaca and London},
  \bibinfo{year}{1979}).

\bibitem[{\citenamefont{Tricomi}(1967)}]{TricomiIntegralEquations}
\bibinfo{author}{\bibfnamefont{F.~G.} \bibnamefont{Tricomi}},
  \emph{\bibinfo{title}{Integral Equations}} (\bibinfo{publisher}{Interscience:
  New York}, \bibinfo{year}{1967}).

\bibitem[{Mat({\natexlab{a}})}]{MathMeijerGFunction}
\eprint{http://mathworld.wolfram.com/MeijerG-Function.html}.

\bibitem[{\citenamefont{Gradshteyn and Ryzhik}(2000)}]{GradshteynRyzhik}
\bibinfo{author}{\bibfnamefont{I.~S.} \bibnamefont{Gradshteyn}}
  \bibnamefont{and} \bibinfo{author}{\bibfnamefont{I.~M.}
  \bibnamefont{Ryzhik}}, \emph{\bibinfo{title}{Tables of Integrals, Series, and
  Products, 6th ed.}} (\bibinfo{publisher}{Academic Press: San Diego, CA},
  \bibinfo{year}{2000}).

\bibitem[{Mat({\natexlab{b}})}]{MathHarmonicNumber}
\eprint{http://mathworld.wolfram.com/HarmonicNumber.html}.

\bibitem[{\citenamefont{Daoud and Cotton}(1982)}]{DaoudCotton1982}
\bibinfo{author}{\bibfnamefont{M.}~\bibnamefont{Daoud}} \bibnamefont{and}
  \bibinfo{author}{\bibfnamefont{J.~P.} \bibnamefont{Cotton}},
  \bibinfo{journal}{J. Phys. (Paris)} \textbf{\bibinfo{volume}{43}},
  \bibinfo{pages}{531} (\bibinfo{year}{1982}).

\bibitem[{\citenamefont{Birshtein and Zhulina}(1984)}]{BirshteinZhulina1984}
\bibinfo{author}{\bibfnamefont{T.~M.} \bibnamefont{Birshtein}}
  \bibnamefont{and} \bibinfo{author}{\bibfnamefont{E.~B.}
  \bibnamefont{Zhulina}}, \bibinfo{journal}{Polymer}
  \textbf{\bibinfo{volume}{25}}, \bibinfo{pages}{1453} (\bibinfo{year}{1984}).

\end{thebibliography}

\end{document}